\documentclass[12pt]{article}
\usepackage{amsmath,amssymb,color}
\usepackage{cite,epsf}

\makeatletter \@addtoreset{equation}{section} \makeatother

\newcommand{\be}{\begin{equation}}
\newcommand{\ee}{\end{equation}}
\newcommand{\bea}{\begin{eqnarray}}
\newcommand{\eea}{\end{eqnarray}}
\def\ie{{\it i.e.}}
\def\del{\partial}
\def\til{\tilde}
\def\a{{\alpha}}
\def\b{{\beta}}
\def\g{{\gamma}}
\def\tg{{\til\g}}
\def\G{{\Gamma}}
\def\Gof{{\G[{\textstyle{1\over4}}]}}
\def\Gtf{{\G[{\textstyle{3\over4}}]}}
\def\d{{\delta}}
\def\e{{\epsilon}}

\def\l{{\lambda}}

\def\ym{{\rm YM}}

\begin{document}

\begin{flushright}
MIFP-06-32 \\
{\bf hep-th/0612157}\\
December\  2006
\end{flushright}

\vspace{10pt}

\begin{center}

{\Large\bf Spacelike strings and jet quenching
\\ [3mm] from a Wilson loop}

\vspace{30pt}

{\large Philip C. Argyres$^1$, Mohammad Edalati$^1$,\\ [2mm] 
and Justin F. V\'azquez-Poritz$^2$}

\vspace{20pt}

{\it $^1$Department of Physics\\ University of Cincinnati, Cincinnati OH 45221-0011}\\ [4mm]
{\it $^2$George P. \& Cynthia W. Mitchell Institute for Fundamental Physics\\
Texas A \& M University, College Station TX 77843-4242}

\vspace{10pt}

\centerline{\tt argyres,edalati@physics.uc.edu, jporitz@physics.tamu.edu}

\vspace{20pt}

\centerline{{\bf{Abstract}}}
\end{center}

\noindent We investigate stationary string solutions with spacelike 
worldsheet in a five-dimensional AdS black hole background, and find 
that there are many branches of such solutions.  Using a non-perturbative 
definition of the jet quenching parameter proposed by Liu {\it et.\ al.}, 
hep-ph/0605178, we take the lightlike limit of these solutions to evaluate 
the jet quenching parameter in an ${\cal N}=4$ super Yang-Mills thermal 
bath.  We show that this proposed definition gives zero jet quenching 
parameter, independent of how the lightlike limit is taken.  In particular, 
the minimum-action solution giving the dominant contribution to the Wilson 
loop has a leading behavior that is linear, rather than quadratic, in the 
quark separation. 

\newpage

\section{Introduction, summary and conclusions}

Results coming from RHIC have raised the issue of how to
calculate transport properties of ultra-relativistic partons in a
strongly coupled gauge theory plasma.   For example, one would like
to calculate the friction coefficient and jet quenching parameter, 
which are measures of the rate at which partons lose energy to the 
surrounding plasma \cite{baier,bsz0002,w0005,kw0106,gvwz0302,kw0304,
s0312,s0405,jw0405}.  With conventional quantum field theoretic tools, 
one can calculate these parameters only when the partons are interacting 
perturbatively with the surrounding plasma.  The AdS/CFT 
correspondence \cite{agmoo9905} may be a suitable
framework in which to study strongly coupled QCD-like plasmas.
Attempts to use the AdS/CFT correspondence to
calculate these quantities have been made in 
\cite{lrw0605,hkkky0605,gub0605,st0605} 
and were generalized in various ways in
\cite{b0605,h0605,cg0605,fgm0605,vp0605,sz0606,cg0606,
lm0606,as0606,psz0606,aem0606,gxz0606,lrw0607,fgmp0607,cgg0607,mtw0607,
cno0607,aev0608,nsw0608,asz0609,fgmp0609,t0610,fgmp0611,cg0611,gxz0611,
a0611,g0611,lrw0612,gubser0612}.

The most-studied example of the AdS/CFT correspondence is that of
the large $N$, large 't Hooft coupling limit of 
four-dimensional ${\cal N}=4$ $SU(N)$ super Yang-Mills (SYM) theory and 
type IIB supergravity on $AdS_5\times S^5$.  
At finite temperature, this SYM theory is equivalent to 
type IIB supergravity on the background of the near-horizon region of 
a large number $N$ of non-extremal D3-branes. From 
the perspective of five-dimensional gauged supergravity, this is the 
background of a neutral AdS black hole whose Hawking temperature equals the 
temperature of the gauge theory \cite{wit9802}.
Since at finite temperature the superconformal invariance
of this theory is broken, and since fundamental matter can be
added by introducing D7-branes \cite{kk0205}, it is thought that
this model may shed light on certain aspects of strongly coupled
QCD plasmas.  

According to the AdS/CFT dictionary, the endpoints of open strings 
on this background can correspond to quarks and antiquarks in the 
SYM thermal bath \cite{ry9803,m9803,rty9803,bisy9803}.  For example, 
a stationary single quark can be described by a string that stretches 
from the probe D7-brane to the black hole horizon.  A semi-infinite 
string which drags behind a steadily-moving endpoint and asymptotically 
approaches the horizon has been proposed as the configuration dual to 
a steadily-moving quark in the ${\cal N}=4$ plasma, and was used to 
calculate the drag force on the quark \cite{hkkky0605,gub0605}.  A 
quark-antiquark pair or ``meson", on the other hand, corresponds to a 
string with both endpoints ending on the D7-brane.  The static limit 
of this string solution has been used to calculate the inter-quark 
potential in SYM plasmas \cite{rty9803,bisy9803}.  Smooth, stationary 
solutions for steadily-moving quark-antiquark pairs exist 
\cite{lrw0607,cgg0607,cno0607,aev0608,asz0609,fgmp0609,lrw0612} but are 
not unique and do not ``drag" behind the string endpoints as in the 
single quark configuration.  This lack of drag has been interpreted 
to mean that color-singlet states such as mesons are invisible to the 
SYM plasma and experience no drag (to leading order in large $N$) even 
though the string shape is dependent on the velocity of the meson with 
respect to the plasma.  Nevertheless, a particular no-drag string 
configuration with spacelike worldsheet \cite{lrw0607,lrw0612} has 
been used to evaluate a lightlike Wilson loop in the field theory 
\cite{lrw0605,b0605,cg0605,vp0605,cg0606,lm0606,as0606,aem0606,
cgg0607,nsw0608,gxz0611}.  It has been proposed that this Wilson loop can 
be used for a non-perturbative definition of the jet quenching parameter 
$\hat q$ \cite{lrw0605}.

The purpose of this paper is to do a detailed analysis of the 
evaluation of this Wilson loop using no-drag spacelike string 
configurations in the simplest case of finite-temperature ${\cal N}=4$ 
$SU(N)$ SYM theory.

\paragraph{Summary.}

We use the Nambu-Goto action to describe the classical dynamics 
of a smooth stationary string in the background of a five-dimensional 
AdS black hole. We put the endpoints of the string on a probe D7-brane 
with boundary conditions which describe a quark-antiquark pair 
with constant separation moving with constant velocity either 
perpendicular or parallel to their separation. 

In section 2, we present the string embeddings describing smooth and 
stationary quark-antiquark configurations, and we derive their equations 
of motion. 
In section 3, we discuss spacelike solutions of these equations. 
We find that there can be an infinite number of spacelike 
solutions for given boundary conditions, although there is always 
a minimum-length solution.

In section 4, we apply these solutions to the calculation of the 
lightlike Wilson loop observable proposed by \cite{lrw0605} to 
calculate the jet quenching parameter $\hat q$, by taking the 
lightlike limit of spacelike string worldsheets \cite{lrw0607,lrw0612}.
We discuss the ambiguities in the evaluation of this Wilson
loop engendered by how the lightlike limit is taken, and by
how self-energy subtractions are performed.  Technical aspects
of the calculations needed in section 4 are collected in two
appendices.  We also do the calculation for Euclidean-signature
strings for the purpose of comparison.

\paragraph{Conclusions.}

We find that the lightlike limit of the spacelike string configuration 
used in \cite{lrw0605,lrw0612} to calculate the jet quenching parameter 
$\hat q$ is not the solution with minimum action for given boundary 
conditions, and therefore gives an exponentially suppressed contribution 
to the path integral.  Regardless of how the lightlike limit is taken, 
the minimum-action solution giving the dominant contribution to the 
Wilson loop has a leading behavior that is linear in its width, $L$.  
Quadratic behavior in $L$ is associated with radiative energy loss by 
gluons in perturbative QCD, and the coefficient of the $L^2$ term is 
taken as the definition of the jet-quenching parameter $\hat q$ 
\cite{lrw0605}.  In the strongly coupled ${\cal N}=4$ SYM theory in 
which we are computing, we find $\hat q=0$.


We now discuss a few technical issues related to the validity of
the dominant spacelike string solution which gives rise to the linear
behavior in $L$. 

Depending on whether the velocity parameter approaches unity from above 
or below, the minimum-action string lies below (``down string") or above 
(``up string") the probe D7-brane, respectively.  The down string 
worldsheet is spacelike regardless of the region of the bulk space in 
which it lies.  On the other hand, in order for the up string worldsheet 
to be spacelike, it must lie within a region bounded by a certain maximum 
radius which is related to the position of the black hole horizon.  The 
lightlike limit of the up string involves taking the maximal radius and 
the radius of the string endpoints to infinity simultaneously, such that 
the string always lies within the maximal radius.  Therefore, even though 
the string is getting far from the black hole, its dynamics are still 
sensitive to the black hole through this maximal radius. 

In the lightlike limit, the up and down strings with minimal action 
both approach a straight string connecting the two endpoints.  This is 
the ``trivial" solution discarded in \cite{lrw0605}, though we do not 
find a compelling physical or mathematical reason for doing so.  If 
the D7-brane radius were regarded as a UV cut-off, then one might 
presume that the dominant up string solution should be discarded, since 
it probes the region above the cut-off.  However, this is unconvincing 
for two reasons.  First, if one approaches the lightlike limit from 
$v>1$, then the dominant solution is a down string, and so evades this 
objection.  Second, and more fundamentally, in a model which treats the 
D7-brane radius as a cut-off one does not know how to compute accurately 
in the AdS/CFT correspondence.  For this reason we deal only with the 
${\cal N}=4$ SYM theory and a probe brane D7-brane, for which the 
AdS/CFT correspondence is precise.

A spacelike string lying straight along a constant radius is discussed 
briefly in \cite{lrw0612}.  This string also approaches the ``trivial" 
lightlike solution in \cite{lrw0605} as the radius is taken to infinity.  
As pointed out in \cite{lrw0612}, this straight string at finite radius
is not a solution of the (full, second order) equations of motion, and 
should be rejected.  We emphasize that our dominant string solutions
are \emph{not} this straight string, even though they approach the
straight string as the D7-brane radius goes to infinity, and are
genuine solutions to the full equations of motion. 

To conclude, the results in this paper show these solutions to be robust, 
in the sense that they give the same contribution to the path integral 
independently of how the lightlike limit is taken.  Furthermore, though 
not a compelling argument, the fact that these solutions do not exhibit 
any drag is consistent with the fact that they give $\hat q=0$.  Therefore, 
for the non-perturbative definition of $\hat q$ given in \cite{lrw0605}, 
direct computation of $\hat q\neq0$ within the AdS/CFT correspondence for 
${\cal N}=4$ $SU(N)$ SYM would require either a compelling argument for 
discarding the leading contribution to the path integral, or a different 
class of string solutions giving the dominant contribution.  On the other 
hand, this computation may simply imply that at large $N$ and strong 't 
Hooft coupling, the mechanism for relativistic parton energy loss in the 
SYM thermal bath gives a linear rather than quadratic dependence on the 
Wilson loop width $L$.

\section{String embeddings and equations of motion}

We consider a smooth and stationary string in the background 
of a five-dimensional AdS black hole with the metric
\be\label{metric}
ds_5^2=h_{\mu\nu} dx^{\mu} dx^{\nu}
=-{r^4-r_0^4\over r^2 R^2}\,dx_0^2+\frac{r^2}{R^2}
(dx_1^2+dx_2^2+dx_3^2)+ {r^2 R^2\over r^4-r_0^4} \,dr^2. 
\ee 
$R$ is the curvature radius of the AdS space, and the black hole 
horizon is located at $r=r_0$.  We put the endpoints of the string 
at the minimal radius $r_7$ that is reached by a probe D7-brane.  
The classical dynamics of the string in this background is described 
by the Nambu-Goto action 
\be\label{ngaction}
S = -\frac{1}{2\pi\a'} \int\! d\sigma d\tau\, \sqrt{-G},
\ee
with
\bea
G &=& \det[h_{\mu\nu}(\del X^\mu/\del\xi^\a) (\del X^\nu/
\del\xi^\b)],
\eea 
where $\xi^\a= \{\tau, \sigma\}$ and $X^\mu=\{x_0,x_1,x_2,x_3,r\}$.

The steady state of a quark-antiquark pair with constant separation 
and moving with constant velocity either perpendicular or parallel 
to the separation of the quarks can be described (up to worldsheet 
reparametrizations), respectively, by the worldsheet embeddings
\bea\label{shape}
{}[v_\perp]: &\quad& 
x_0 = \tau, \quad
x_1= v \tau, \qquad\ \,\,\,
x_2= \sigma, \quad
x_3= 0,\quad
r = r(\sigma), \nonumber\\ [1mm]
{}[v_{||}]: &\quad&
x_0 = \tau, \quad
x_1= v \tau + \sigma, \quad
x_2= 0, \quad
x_3= 0,\quad
r = r(\sigma).
\eea
For both cases, we take boundary conditions
\be\label{bcs} 
0\le \tau\le T,\qquad
-L/2 \le\sigma\le L/2, \qquad
r(\pm L/2)= r_7,
\ee 
where $r(\sigma)$ is a smooth embedding.

The endpoints of strings on D-branes satisfy Neumann boundary conditions
in the directions along the D-brane, whereas the above boundary conditions
are Dirichlet, constraining the string endpoints to lie along fixed
worldlines a distance $L$ apart on the D7-brane.  The correct way to 
impose these boundary conditions is to turn on a worldvolume background 
$U(1)$ field strength on the D7-brane \cite{hkkky0605} to keep the string endpoints a distance $L$ apart.  Thus at finite $r_7$, it is physically
more sensible to describe string solutions for a fixed force on the 
endpoints instead of a fixed endpoint separation $L$.\footnote{We thank 
A. Karch for discussions on this point.}  Our discussion of spacelike 
string solutions in the next section will describe both the 
force-dependence and the $L$-dependence of our solutions.  In the 
application to evaluating a Wilson loop in section 4, though, we are 
interested in string solutions (in the $r_7\to\infty$ limit) with 
endpoints lying along the given loop, \ie, at fixed $L$.

According to the AdS/CFT correspondence, strings ending on the 
D7-brane are equivalent to quarks in a thermal bath in 
four-dimensional finite-temperature ${\cal N}{=}4$ $SU(N)$ super 
Yang-Mills (SYM) theory. The standard gauge/gravity dictionary is 
that $N=R^4/(4\pi\a'^2 g_s)$ and $\l = R^4/\a'^2$  
where $g_s$ is the string coupling, $\l:=g_\ym^2 N$ is the 
't Hooft coupling of the SYM theory.
In the semiclassical string limit, {\it i.e.}, 
$g_s\to0$ and $N\to\infty$, the supergravity approximation in the 
gauge/gravity correspondence holds when the curvatures are much 
greater than the string length $\ell_s := \sqrt{\a'}$.  
Furthermore, in this limit, one identifies
\be\label{dictionary}
\b = \pi R^2/r_0, \qquad
m_0 = r_7/(2\pi\a') ,
\ee 
where $\b$ is the (inverse) temperature of the SYM thermal bath, and 
$m_0$ is the quark mass at zero temperature.   

\begin{figure}[t]
   \epsfxsize=3.0in \centerline{\epsffile{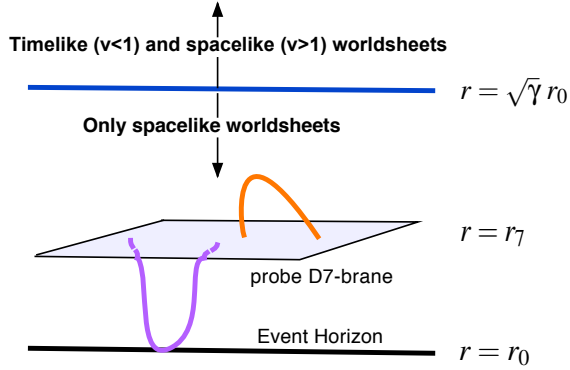}}
   \caption[FIG. \arabic{figure}.]{\footnotesize{Both timelike and 
spacelike worldsheets can exist above the radius $r=\sqrt{\gamma} r_0$ 
(blue line) for $v<1$ and $v>1$, respectively. On the other hand, 
only spacelike worldsheets exist in the region between the blue 
line and the event horizon, given by $r_0<r<\sqrt{\gamma} r_0$.}}
\label{fig6}
\end{figure}

It will be important to note that the velocity parameter $v$ entering
in the string worldsheet embeddings (\ref{shape}) is \emph{not}
the proper velocity of the string endpoints.  Indeed, from (\ref{metric})
it is easy to compute that the string endpoints at $r=r_7$
move with proper velocity
\be\label{realV}
V = {r_7^2\over\sqrt{r_7^4-r_0^4}}\, v.
\ee
We will see shortly that real string solutions must have the
same signature everywhere on the worldsheet.  Thus a string
wroldsheet will be timelike or spacelike depending on whether
$V$, rather than $v$, is greater or less than 1.  Thus, translating
$V\lessgtr 1$ into corresponding inequalities for the velocity 
parameter $v$, we have
\bea\label{spacetime}
\begin{array}{c}
\mbox{timelike}\\
\mbox{string worldsheet}\\
\end{array} 
&\Leftrightarrow &
\quad\, \mbox{both}\quad\ v<1\ \,(\g^2>1) \ \mbox{and}\ 
z_7 > \sqrt\g ,\nonumber\\
&&\\
\begin{array}{c}
\mbox{spacelike}\\
\mbox{string worldsheet}\\
\end{array} 
&\Leftrightarrow &
\left\{
\begin{array}{l}
\mbox{either}\ \ \,v\ge1\ \,(\g^2<0)\ \mbox{and any}\ z_7, \\ [2mm]
\mbox{or}\qquad\ \, v<  1\ \,(\g^2>1)\ \mbox{and}\  
z_7  < \sqrt\g .\\
\end{array}
\right.\nonumber
\eea
Here we have defined the dimensionless ratio of the D7-brane 
radial position to the horizon radius,
\be
z_7 := {r_7\over r_0},\qquad\mbox{and}\qquad
\g^2 := {1\over 1-v^2 } .
\ee
Furthermore, since the worldsheet has the same signature everywhere, 
this implies that timelike strings can only exist for $r>\sqrt\g\, 
r_0$, but spacelike strings may exist at all $r$, as illustrated in 
figure \ref{fig6}.  In this respect, $r=\sqrt\g\,r_0$ plays a role 
analogous to that of the ergosphere of a Kerr black hole, although 
in this case it is not actually an intrinsic feature of the background 
geometry but instead a property of certain string configurations 
(\ref{shape}) in the background geometry (\ref{metric}).

With the embeddings (\ref{shape}) and boundary conditions
(\ref{bcs}), the string action becomes\footnote{These expressions
for the string action are good only when there 
is a single turning point around which the string is symmetric. 
We will later see that for $[v_\perp]$ there exist solutions 
with multiple turns. For such solutions the limits of integration 
in (\ref{singleturnaction}) are changed, and appropriate terms for 
each turn of the string are summed.}
\bea\label{singleturnaction}
{}[v_\perp]: &\quad& 
S=\frac{- T}{\g\,\pi\a'}
\int_0^{L/2} d\sigma \sqrt{{r^4-\g^2 r_0^4\over R^4}
+ {r^4-\g^2 r_0^4\over r^4-r_0^4} r'^2}, \nonumber\\
{}[v_{||}]: &\quad&
S=\frac{- T}{\g\,\pi\a'}
\int_0^{L/2} d\sigma \sqrt{\g^2 {r^4-r_0^4\over R^4}
+ {r^4-\g^2 r_0^4\over r^4-r_0^4} r'^2}, 
\eea
where $r' := \del r/\del\sigma$.  The resulting equations of motion are
\bea\label{E} 
{}[v_\perp]: &\quad& 
r'^2=\frac{1}{\g^2\,a^2 r_0^4 R^4}
(r^4-r_0^4) (r^4-\g^2 [1+a^2] r_0^4) , \nonumber\\
{}[v_{||}]: &\quad&
r'^2=\frac{\g^2}{a^2 r_0^4 R^4}
(r^4-r_0^4)^2 {(r^4-[1+a^2]r_0^4) \over (r^4-\g^2 r_0^4)} , 
\eea
where $a^2$ is a real integration constant.  Here we have taken
the first integral of the second order equations of motion
which follows from the existence of a conserved momentum in 
the direction along the separation of the string endpoints. 
Since $a$ is associated with this conserved momentum, $|a|$ is 
proportional to the force applied (via a constant background
$U(1)$ field strength on the D7 brane) to the string endpoints
in this direction \cite{hkkky0605}.  

Although we have written $a^2$ as a square, it can be either 
positive or negative.  Using (\ref{E}), the determinant of the induced 
worldsheet metric can be written as
\bea\label{G}
{}[v_\perp]: &\quad& 
G=-\frac{1}{\g^4\,a^2 r_0^4 R^4}
(r^4-\g^2 r_0^4)^2, \nonumber\\
{}[v_{||}]: &\quad&
G=-\frac{1}{a^2 r_0^4 R^4} (r^4-r_0^4)^2 .
\eea
Thus, the sign of $G$ is the same as that of $-a^2$ (since
the other factors are squares of real quantities).
In particular, the worldsheet is timelike ($G<0$) for $a^2>0$ and
spacelike ($G>0$) for $a^2<0$.

The reality of $r'$ implies that the right sides of (\ref{E})
must be positive in all these different cases, which implies 
certain allowed ranges of $r$.  Therefore, there can only be 
real string solutions when the ends of the string, at $r=r_7$, 
are within this range.  The edges of this range are (typically) 
the possible turning points $r_t$ for the string, whose possible 
values will be analyzed in the next
section.

Given these turning points, (\ref{E}) can be integrated to give
\bea\label{L}
{}[v_\perp]: &\quad& 
{L\over\b}=\frac{2|a\g|}{\pi} \left|\int_{z_t}^{z_7} 
{dz\over\sqrt{(z^4-1)(z^4-\g^2[1+a^2])}}\right|, \nonumber\\
{}[v_{||}]: &\quad&
{L\over\b}=\frac{2|a|}{\pi|\g|} \left|\int_{z_t}^{z_7} 
{dz\sqrt{z^4-\g^2}\over(z^4-1)\sqrt{z^4-[1+a^2]}}\right|,
\eea
where we have used $r_0=\pi R^2/\b$. Also, in (\ref{L}) we have
rescaled $z=r/r_0$ and likewise $z_t:=r_t/r_0$ and $z_7:=r_7/r_0$.
(The absolute value takes care of cases where $z_7<z_t$.)  
These integral expressions determine the integration constant
$a^2$ in terms of $L/\b$ and $v$. 

Also, we can evaluate the action for the solutions of (\ref{E}):
\bea\label{S}
{}[v_\perp]: &\quad& 
S= \pm\frac{T\sqrt\l}{\g\b}
\int_{z_t}^{z_7} {(z^4-\g^2)\, dz \over
\sqrt{(z^4-1)(z^4-\g^2[1+a^2])}} 
, \nonumber\\
{}[v_{||}]: &\quad&
S= \pm\frac{T\sqrt\l}{\g\b}
\int_{z_t}^{z_7} dz \sqrt{z^4-\g^2 \over
z^4-[1+a^2]} ,
\eea
where we have used $R^2/\a'=\sqrt\l$.
The plus or minus signs are to be chosen depending on the
relative sizes of $z_7$, $z_t$, and $\g^2$, and will
be discussed in specific cases below.
For finite $z_7$, these integrals are convergent.  They diverge
when $z_7\to\infty$ and need to be regularized by subtracting
the self-energy of the quark and the antiquark \cite{ry9803,m9803}, 
which will be discussed in more detail in section 4.

Note that, in writing (\ref{L}) and (\ref{S}), we have 
assumed that the string goes from $z_7$ to the turning
point $z_t$ and back only once.  We will see that more 
complicated solutions with multiple turning points are 
possible. For these cases, one must simply add an appropriate 
term, as in (\ref{L}) and (\ref{S}), for each turn of the string.

\section{Spacelike solutions}

Positivity of the determinant of the induced worldsheet
metric (\ref{G}) implies that the integration constant $a^2<0$ 
for spacelike configurations.  It is convenient to define
a real integration constant $\a$ by
\be
\a^2 := -a^2 >0.
\ee
As remorked above, $\a$ is proportional to the magnitude of a
background $U(1)$ field strength on the D7 brane.
We will now classify the allowed ranges of $r$ for which $r'^2$ 
is positive in the equations of motion (\ref{E}).  These ranges, 
as well as the associated possible turning points of the string 
depend on the relative values of $\a$, $v$ and 1.

\subsection{Perpendicular velocity}

The configurations of main interest to us are those for which 
the string endpoints move in a direction perpendicular to their 
separation.  As we will now see, the resulting solutions have 
markedly different behavior depending on whether the velocity 
parameter is greater or less than 1.  

\subsubsection{$\sqrt{1-z_7^{-4}}<v<1$}

If $v<1$, we have seen that the string worldsheet can
be spacelike as long as $v>\sqrt{1-z_7^{-4}}$.
A case-by-case classification of the possible
turning points of the $v_\perp$ equation in (\ref{E}) gives the
following table of possibilities:
\medskip

\centerline{{\footnotesize
\begin{tabular}{l|rlrl} 
\normalsize parameters&\multicolumn{4}{c}{\normalsize allowed ranges}\\ \hline
$0<\a<v<1$\ \ &&&$1\le$&$\!\!\!\!z^4\le\g^2(1-\a^2)$\\
$0<v<\a<1$&$\g^2(1-\a^2)\le$&$\!\!\!\!z^4\le1$&&\\
$0<v<1<\a$&$0\le$&$\!\!\!\!z^4\le1$&&\\
\end{tabular}
}}\medskip

\noindent The left column 
of allowed ranges are those that lie inside the horizon and 
the right one are the allowed ranges outside the horizon.  

At the horizon, $r'=0$ and the string becomes tangent to $z=1$ 
at finite transverse distance giving a smooth turning point for 
the string. In the last entry in the above table, ``$0\le z^4$" 
indicates that there is no turning point before meeting the 
singularity at $z=0$ and the string necessarily meets the 
singularity.

Since only string solutions that extend into the $z>1$ region can 
reliably describe quarks, this eliminates the left column of allowed 
ranges.  Thus, the only viable configurations are those in the right 
column with $\a<v$, which all have turning points at $1$ or 
$\g^2(1-\a^2)$.  This restricts the D7-brane minimum radius to lie 
between these two turning points which, in turn, gives rise to string 
configurations with multiple turns.

\begin{figure}[t]
   \epsfxsize=5.4in \centerline{\epsffile{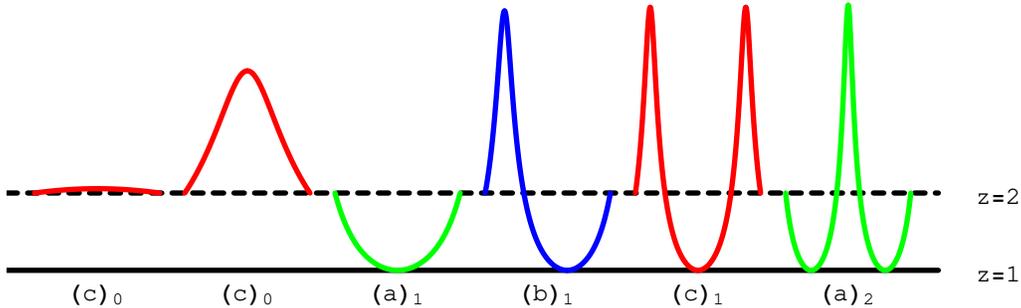}}
   \caption[FIG. \arabic{figure}.]{\footnotesize{Spacelike
   string solutions with fixed $L/\b=0.25$, $\g=20$
   ($v\approx0.99875$), $z_7=2$, and with low values of $n$ 
   (the number of turns at the horizon).  The horizon is the
   solid line at $z=1$, and the minimum radius of the D7-brane
   is the dashed line at $z=2$.}}\label{fig2}
\end{figure}

In order for the D7-brane to be within the allowed
range $1<z_7^4<\g^2(1-\a^2)$,
the parameter $v$ must be at least $v^2 > 1-z_7^{-4}$.
For a given $z_7$, $v$, and $\a$ satisfying
these inequalities, we integrate (\ref{L}) to obtain $L/\b$. 
There are two choices for the range of integration: $[1,z_7]$ 
and $[z_7,\g^2(1-\a^2)]$.  The first one is appropriate for 
a string which decends down to the horizon and then turns 
back up to the D7-brane; we will call this a ``down string".  
The second range describes a string which ascends to larger 
radius and then turns back down to the D7-brane; we will call 
this an ``up string".  Given these two behaviors, it is clear 
that we can equally well construct infinitely many other
solutions by simply alternating segments of up and down
strings.  In particular, there are three possible series
of string configurations, which we will call the (a)$_n$,
(b)$_n$, and (c)$_n$ series.  An (a)$_n$ string starts with
a down string then adds $n-1$ pairs of up and down strings, 
thus ending with a down string; a (b)$_n$ string concatenates $n$
pairs of up and down strings--- for example,  starts with an up string
and ends with a down string; and a (c)$_n$ string starts with
an up string and then adds $n$ pairs of down and up strings, thus
ending with an up string.  $n$ counts the number of turns the string makes at the
horizon, $z=1$. In particular, for the (a)$_n$ and 
(b)$_n$ series, $n$ is an integer $n\ge1$, while for the (c)$_n$ series, 
$n\ge0$. Examples of these string configurations appear 
in Figure \ref{fig2}. If the separation of the ends of the up and
down strings are $L_{\rm up}$ and $L_{\rm down}$, respectively,
then the possible total separations of the strings fall
into three classes of lengths
\bea
L_{{\rm (a)},n} &=& n L_{\rm down} + (n-1) L_{\rm up} ,\nonumber\\
L_{{\rm (b)},n} &=& n L_{\rm down} + n L_{\rm up} ,\nonumber\\
L_{{\rm (c)},n} &=& n L_{\rm down} + (n+1) L_{\rm up} .
\eea
 
\begin{figure}[t]
   \epsfxsize=5.8in \centerline{\epsffile{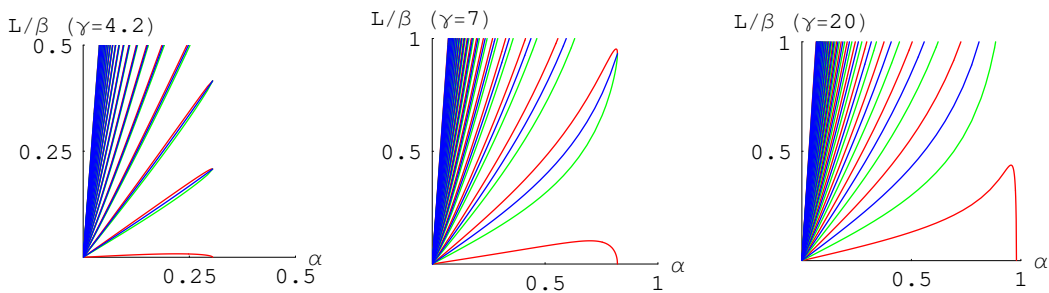}}
   \caption[FIG. \arabic{figure}.]{\footnotesize{$L/\b$ as a 
   function of $\a$ for spacelike string configurations 
   with perpendicular velocity, $z_7=2$ and $\g=4.2$, $7$, and $10$.  
   Green curves correspond to the (a)--series, blue to (b)--series, 
   and red to (c)--series.  Only the series up to $n=20$ are shown; 
   the rest would fill the empty wedge near the $L/\b$ axis. 
   Note that the scale of the $\g=4.2$ plot is half that of the 
   other two.}}
   \label{fig1}
\end{figure}

Figure \ref{fig1} illustrates the systematics of the $L_{{\rm (a,b,c)},n}$ 
dependence on $\a$.  Here we have chosen $z_7=2$, so the
minimum value of $v$ for the solutions to exist has $\g=4$.
The leftmost plot illustrates that, for small $\g$, 
$L_{{\rm (c)},0}=L_{\rm up} \ll L_{\rm down}$ for all $\a$.
Thus, for each $n\ge1$, $L_{{\rm (a)},n}\approx L_{{\rm (b)},n}
\approx L_{{\rm (c)},n}$, and are virtually indistinguishable
in the figure.  As $\g$ increases, $L_{\rm up}$ and
$L_{\rm down}$ begin to approach each other for most $\a$,
except for $\a$ near $\a_{\rm max}:=\sqrt{1-z_7^4/\g^2}$, 
where $L_{\rm up}$ decreases sharply back to zero.

This behavior implies that, for every fixed $L$ and $v$, there
is a very large number of solutions\footnote{Although $n$ does not 
formally have an upper bound, as $n$ increases the turns of the 
string become sharper and denser. Therefore, for large enough $n$ the
one can no longer ignore the backreaction of the string on the 
background.} in each series but that the minimum value of $n$ 
that occurs decreases as $v$ increases. In detail, it is not too 
hard to show that the pattern of appearance of solutions as $v$ 
increases for fixed $L$ is as follows: if for a given $v$ there 
is one solution ({\it i.e.}, value of $\a$) for each (a,b,c)$_n$--string 
with $n> n_0$, then as $v$ increases first {\it two} (c)$_{n_0}$ 
solutions will appear, then the (c)$_{n_0}$ solution with the 
greater $\a$ will disappear just as a (b)$_{n_0}$ and an (a)$_{n_0}$ 
solution appear.  Also, $\a({\rm (a)}_n)<\a({\rm (b)}_n)<\a({\rm (c)}_n)$. 
For example, in Figure \ref{fig1}, when $L/\b=0.25$ and $\g=4.2$,
there are (a,b,c)$_n$ solutions for $n\ge2$.  Increasing $v$ to
$\g=7$ (for the same $L$), there are now (a,b,c)$_n$ solutions for
$n\ge1$.  Increasing $v$ further to $\g=20$, there are now in 
addition two (c)$_0$ ({\it i.e.}, up string) solutions.
Figure \ref{fig2} plots the string solutions when $z_7=2$, $L/\b=0.25$,
and $\g=20$, for low values of $n$.

Note that if one keeps the D7-brane $U(1)$ field strength, $\a$,
constant instead of the endpoint separation, $L$, then there
will still be an infinite sequence of string solutions 
qualitatively similar to that shown in Figure 2.  In this case 
the endpoint separation $L$ increases with the number of
turns.

\begin{figure}[t]
   \epsfxsize=5.7in \centerline{\epsffile{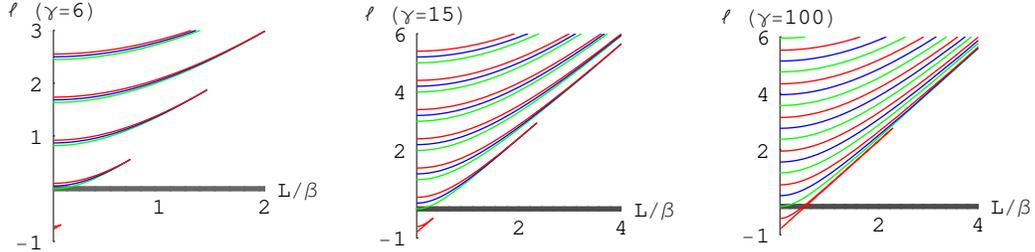}}
   \caption[FIG. \arabic{figure}.]{\footnotesize{Spacelike
   string lengths $\ell$ in units of 
   $\sqrt\l/\b$ as a function of endpoint separation 
   $L/\b$ and $z_7=2$, for $\g=6,15$ and $100$.
   The gray line along the $L/\b$ axis is the (subtracted)
   length of a pair of straight strings stretched between the
   D7-brane and the horizon. Note that the scale of the
   $\g=6$ plot is half that of the others.}}\label{fig3}
\end{figure}

The action for spacelike configurations is
imaginary because the Nambu-Goto Lagrangian is $\sqrt{-G}
=\pm i \sqrt{G}$.  Ignoring the $\pm i$ factor (which
we will return to in the next section), the integral
of $\sqrt{G}$ just gives the area of the worldsheet.
Dividing by the ``time" parameter $T$ in (\ref{S}) then
gives the length of the string: $\ell=\pm i S/T$.
Figure \ref{fig3} plots the lengths of the various series of
string configurations for increasing values of
the velocity parameter.  There are negative lengths
because the length of a pair of straight strings stretched
between the D7-brane and the horizon has been subtracted,
for comparison purposes.
It is clear from the figure that the (c)$_0$ up strings 
are the shortest for any given $L$ less than a 
velocity-dependent critical value.  Furthermore, 
for $L$ small enough, they are also shorter
than the straight strings.  

In particular, the shorter (larger $\a$) of the two up 
strings has the smallest $\ell$ of all. As $v\to1$, the 
critical value of $L$ below which the up string is the 
solution with the minimum action increases without bound. 
In this case, any of the other spacelike strings will 
decay to this minimum-action configuration. Therefore, 
it is this configuration which must be used for any 
calculations of physical quantities, such as the jet 
quenching parameter $\hat q$.

\subsubsection{$v>1$}

A case-by-case classification of the possible
turning points of the $v_\perp$ equation in (\ref{E}) 
when $v>1$ gives the
following table of possibilities:
\medskip

\centerline{{\footnotesize
\begin{tabular}{l|rlrl} 
\normalsize parameters&\multicolumn{4}{c}{\normalsize allowed ranges}\\ \hline
$0<\a<1<v$&&&$1\le$&$\!\!\!\!z^4<\infty$\\ 
$0<1<\a<v$&$0\le$&$\!\!\!\!z^4\le\g^2(1-\a^2)$&$1\le$&$\!\!\!\!z^4<\infty$\\ 
$0<1<v<\a$&$0\le$&$\!\!\!\!z^4\le1$&$\g^2(1-\a^2)\le$&$\!\!\!\!z^4<\infty$\\ 
\end{tabular}
}}\medskip

\noindent The left column 
of allowed ranges are those that lie inside the horizon, while 
the right one lists the allowed ranges which are outside the horizon.  
At the horizon, $r'=0$ and the string becomes tangent 
to $z=1$ at finite transverse distance.  This can be either
a smooth turning point for the string or, if there is an allowed 
region on the other side of the horizon, then the string can have
an inflection point at the horizon and continue through it. 
From the table, we see that this can only happen at the crossover between
the last two lines---in other words, when $\a=v>1$.
In the above table we have written
``$0\le z^4$" when there is no turning point in an
allowed region before the singularity at $z=0$.
In these cases, a string extending towards smaller $z$
will necessarily meet the singularity.
As before, we are only interested in string solutions that extend 
into the $z>1$ region.  This eliminates
the left column of allowed ranges, with the possible
exception of the $v=\a>1$ crossover case, for which the string might
inflect at the horizon and then extend inside.  However, if it does
extend inside, then it will hit the singularity. Therefore, we can also discard this
possibility as being outside the regime of validity of our
approximation.  Thus, the only viable configurations
are those given in the right column, which all have turning 
points at either $1$ or $\g^2(1-\a^2)$, or else go off to infinity.
Since we want to identify the quarks with the ends of the
strings on the D7-brane, we are only interested in 
string configurations that begin and end at $z_7>1$, and
so discard configurations which go off to $z\rightarrow\infty$ instead
of turning.
Thus the $v>1$ ranges compatible with these conditions all 
have only one turning point, describing strings dipping
down from the D7-brane and either turning  
at the horizon or above it, depending on $\a$ versus $v$.   

\begin{figure}[t]
   \epsfxsize=3.0in \centerline{\epsffile{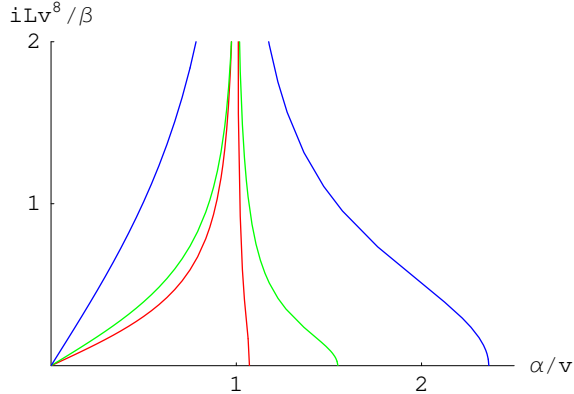}}
   \caption[FIG. \arabic{figure}.]{\footnotesize{$Lv^8/\b$ as a 
   function of $\a/v$ for spacelike string configurations 
   with perpendicular velocity, $z_7=2$, and $v=1.005$ (red), $1.05$
   (green), and $1.2$ (blue).}}
   \label{fig4}
\end{figure}

Indeed, it is straightforward to check
that for any $L$ there are two $v>1$ solutions, one with
$\a>v$ and one with $\a<v$. 
$L/\b$ as a function of $\a$ is plotted in Figure \ref{fig4}.
(The rescalings by powers of $v$ are just so the curves
will nest nicely in the figure.)
 The $\a<v$ solutions are long strings which
turn at the horizon, while the $\a>v$ solutions are
short strings with turning point $z_t^4 = (\a^2-1)/(v^2-1)$.  
The norm of the action for these configurations (which is proportional
to the length of the strings) is likewise greater for the
$\a<v$ solutions than for the $\a>v$ ones.

If, instead, one keeps the D7-brane $U(1)$ field strength $\a$
constant, then there is at most a single string solution with a
given velocity $v$.

\subsection{Parallel velocity}

For completenes, though we will not be using these confurations in
the rest of the paper, we briefly outline the set of solutions
for suspended spacelike
string configurations with velocity parallel to the endpoint 
separation.  A case-by-case analysis of the
equations of motion (\ref{E}) gives the following
table of allowed ranges for string solutions:

\medskip\centerline{{\footnotesize 
\begin{tabular}{l|rl} 
\normalsize parameters &\multicolumn{2}{c}{\normalsize allowed 
ranges}\\ \hline
$0<(\a,v)<1$\ \ & $1-\a^2 \le z^4<1$ & $1<z^4\le\g^2$ \\
$0<v<1<\a$      & $0\le       z^4<1$ & $1<z^4\le\g^2$ \\ \hline
$0<\a<1<v$      & $1-\a^2 \le z^4<1$ & $1<z^4<\infty$ \\ 
$0<1<(\a,v)$    & $0\le       z^4<1$ & $1<z^4<\infty$ \\ 
\end{tabular}}}\medskip

\noindent Although $z=1$ is always included in the allowed ranges,
in the table we have split each range into 
two regions: one inside the horizon and one outside.  The reason 
is that the string equation of motion (\ref{E}) near $r=r_0$
is $r'^2 \sim (r-r_0)^2$, whose solutions are
of the form $r-r_0 \sim \pm e^{\pm \sigma}$.  This implies that these solutions
asymptote to the horizon and never turn.  Thus, the parallel
spacelike strings can never cross the horizon.

As always, we only look at solutions that extend into the $z>1$
region, since that is where we can reliably put D7-branes.  
This eliminates the left-hand column of configurations. 
Recall that the signature of the worldsheet metric for a string 
with $v<1$ changes at $z=\sqrt{\g}$. A string that reaches 
this radius will have a cusp there. This  is qualitatively 
similar to the timelike parallel solutions with cusps described 
in \cite{aev0608}, except that in the spacelike case the strings 
extend away from the horizon (towards greater $z$). Thus, 
the string solutions corresponding to the ranges in the 
right-hand column all either asymptote to $z=1$, go off to 
infinity or have a cusp at $z=\sqrt{\g}$.  The first two 
cases do not give strings with two endpoints on the D7-brane
at $z=z_7$.  Therefore, the only potentially interesting
configurations for our purposes are those with $1<z_7 < z<\sqrt{\g}$,
which occur for $v<1$ and any $\a$.  However, since these 
configurations have cusps, their description in terms of 
the Nambu-Goto action is no longer complete. That is, 
there must be additional boundary conditions specified, 
which govern discontinuities in the first derivatives of 
the string shape. As discussed in \cite{aev0608} for the 
analogous timelike strings, these cusps cannot be avoided 
by extending the string to include a smooth but self-intersecting 
closed loop, since real string solutions cannot change 
their worldsheet signature.

\section{Application to jet quenching}

We will now apply the results of the last two sections
to the computation of the expectation value of a certain 
Wilson loop $W[{\cal C}]$ in the SYM theory.
The interest of this Wilson loop is that it has been
proposed \cite{lrw0605} as a non-perturbative definition 
of the jet quenching parameter $\hat q$. This medium-dependent 
quantity measures the rate per unit distance traveled at which 
the average transverse momentum-squared is lost by a parton 
moving in plasma \cite{baier}.

In particular, \cite{lrw0605} considered a rectangular loop 
$\cal C$ with parallel lightlike edges a distance $L$ apart 
which extend for a time duration $T$.  Motivated by a 
weak-coupling argument, the leading behavior of $W[{\cal C}]$ 
(after self-energy subtractions) for large $T$ and $L\b\ll 1$ 
is claimed to be
\be\label{WL}
\langle W^{A}({\cal C})\rangle =\hbox{exp}[-\frac{1}{4}~{\hat q}~TL^2],
\ee
where $\langle W^{A}({\cal C})\rangle$ is the thermal expectation 
value of the Wilson loop in the adjoint representation. We will 
simply view this as a definition of $\hat q$.\footnote{This differs 
by a constant factor from the definition written in \cite{lrw0605} 
since here it is expressed in the reference frame of the plasma 
rather than that of the parton.} Note that exponentiating the 
Nambu-Goto action gives rise to the thermal expectation value of 
the Wilson loop in the fundamental representation $\langle {W^{F}
({\cal C})}\rangle$. Therefore, we will make use of the relation 
$\langle W^{A}({\cal C})\rangle \approx \langle W^{F}({\cal C})\rangle^2$, 
which is valid at large $N$. 

Self-energy contributions are expected to contribute
on the order of $TL^0$ and, since this is independent of $L$, their
subtraction does not affect the $L$-dependence of the
results.  The subtraction is chosen to remove infinite constant
contributions, but are ambiguous up to finite terms.\footnote{Note
that \cite{drukker} shows that the correct treatment of the Wilson 
loop boundary conditions should automatically and uniquely subtract
divergent contributions;  it would be interesting to evaluate 
our WIlson loop using this prescription instead of the more
{\it ad hoc} one used here and throughout the literature.}
However, there may be other leading contributions of order $TL^{-1}$ 
or $TL$.   For example, as we discussed in the conclusions, a term 
linear in $L$ would be consistent with energy loss by elastic 
scattering.  Therefore, one requires a subtraction prescription.  
We will assume the following one:  extract $2^{-3/2}\hat q$ as the 
coefficient of $L^2$ in a Laurent expansion of the action around
$L=0$.  Thus, concretely,
\be\label{qhat}
W[{\cal C}] \sim \exp\left\{-T
\left( \cdots + {\a_{-1}\over L} + \a_0 + \a_1 L 
+ {\hat q\over4}L^2 +\cdots\right)\right\}.
\ee
Implicit in this is a choice of finite parts of
leading terms to be subtracted, which could affect the
value of $\hat q$; we have no justification for this
prescription beyond its simplicity.
We will see that this issue of $L$-dependent leading
terms indeed arises in the computation of $\hat q$ using the AdS/CFT
correspondence.

There is a second subtlety in the definition of 
$\hat q$ given in (\ref{qhat}), which involves how the lightlike 
limit of $\cal C$ is approached.  In the AdS/CFT correspondence, 
we evaluate the expectation value of the Wilson loop as the 
exponential $\exp\{iS\}$ of the Nambu-Goto action for a string 
with boundary conditions corresponding to the Wilson loop $\cal C$.
If we treat $\cal C$ as the lightlike limit of a
sequence of timelike loops, then the string worldsheet
will be timelike and the exponential will
be oscillatory, instead of exponentially suppressed in $T$
as in (\ref{qhat}).  The exponential suppression requires
either an imaginary action (of the correct sign) or
a Wick rotation to Euclidean signature.  

The authors of \cite{lrw0605} advocate the use of the
lightlike limit of spacelike strings to evaluate
the Wilson loop \cite{lrw0607,lrw0612}.
Below, we will evaluate the Wilson loop using both the
spacelike prescription and the Euclidean one.  Our interest
in the Euclidean Wilson loop is mainly for comparison
purposes and to help elucidate some subtleties in the
calculation; we emphasize that it is \emph{not} the one
proposed by the authors of \cite{lrw0605} to evaluate $\hat q$.  
(Though the Euclidean prescription is the usual one for evaluating
static thermodynamic quantities, we are here evaluating
a non-static property of the SYM plasma and so the usual
prescription may not apply.)

In both cases we will find that, regardless of the manner 
in which the above ambiguities are resolved, the computed value 
of $\hat q$ is zero.

\subsection{Euclidean Wilson loop}

Euclidean string solutions \cite{aev0608} are reviewed in appendix A.  
Here we just note their salient properties.  In Euclidean signature, 
nothing special happens in the limit $V\to1$ ($v\to\sqrt{1-z_7^{-4}}$).
When $V=1$ there are always only two Euclidean string solutions: the 
``long string", with turning point at the horizon $z=1$, and the ``short 
string", with turning point above the horizon.  The one which gives the 
dominant contribution to the path integral is the one with smallest 
Euclidean action.  For endpoint separation $L$ less than a critical value,
the dominant solution is the short string.  This is the string configuration 
that remains the furthest from the black hole horizon \cite{aev0608}.   

We are interested in evaluating the Euclidean string action for the short 
string in the small $L$ limit (the so-called ``dipole approximation").  
However, there is a subtlety associated with taking this limit since it 
does not commute with taking the $z_7\to\infty$ limit, which corresponds to 
infinite quark mass.  Recall that the quark mass scales as $r_7$ in string 
units; introduce a rescaled length parameter
\be
\e := {1\over z_7} = {r_0\over r_7}
\ee
associated with the Compton wavelength of the quark.  Then the behavior of 
the Wilson loop depends on how we parametrically take the $L\to0$ and 
$\e\to0$ limits.  For instance, if one keeps the mass ($\e^{-1}$) fixed 
and takes $L\to0$ first, then the Wilson loop will reflect the overlap of 
the quark wave functions. On the other hand, if one takes $\e\to0$ before 
$L$, then the Wilson loop should reflect the response of the plasma to 
classical sources.  The second limit is presumably the more physically
relevant one for extracting the $\hat q$ parameter.  We perform the 
calculation in both limits in appendix A to verify this intuition.

In the $L\to0$ limit at fixed (small) $\e$, the action of the short string 
as a function of $L$ and $\e$ is found in appendix A to be
\be\label{euc1}
S= {\pi T\sqrt\l\over\sqrt2\b^2}
\left\{ {L\over\e^2} \left[1 +\textstyle{1\over4}\e^4+{\cal O}(\e^8)\right]
- {\pi^2 L^3\over\b^2\e^4} \left[ \textstyle{1\over3}
-{1\over6}\e^4  +{\cal O}(\e^8)\right]
+{\cal O}\left(\textstyle{L^5\over\b^4\e^6}\right)\right\}.
\ee
(In fact, this result is valid as long as $L\to0$ as $L\propto\e$ or faster.) 
The main thing to note about this expression is that it is divergent as 
$\e\to0$.  This is \emph{not} a self-energy divergence that we failed to 
subtract, since any self-energy subtraction ({\it e.g.}, subtracting the 
action of two straight strings extending radially from $z=z_7$ to $z=1$)
will be independent of the quark separation and so cannot cancel the 
divergences in (\ref{euc1}).  (In fact, it inevitably adds an $\e^{-1}L^0$
divergent piece.)  This divergence as the quark mass is taken infinite is 
a signal of the unphysical nature of this order of limits.

The other order of limits, in which $\e\to0$ at fixed (small) $L$, is 
expected to reflect more physical behavior.  Indeed, appendix A gives
\be\label{euc2}
{\b\hat S\over T \sqrt\l} = 
- {0.32}\, {\b\over L} + {1.08} 
- {0.76}\, {L^3\over\b^3} + {\cal O}(L^7) .
\ee
Here $\hat S$ is the action with self-energy subtractions.
This result (which is in the large mass, or $\e\to0$, 
limit) is finite for finite quark separation $L$.  The $L^{-1}$ term 
recovers the expected Coulombic interaction.
Since there is no $L^2$ term in (\ref{euc2}), the subtraction prescription 
(\ref{qhat}) implies that the Euclidean analog of the jet quenching 
parameter vanishes.

For the sake of comparison, we also compute the long string action in 
this limit with the same regularization in Appendix A, giving
\be\label{euc3}
{\hat S_{\rm long}\over T \sqrt\l} = 
+2.39\, {L^2\over\b^3} + {\cal O}(L^4).
\ee
This does have the leading $L^2$ dependence, giving rise to an unambiguous 
nonzero $\hat q$.  But it is exponentially suppressed compared to the short 
string contribution (\ref{euc2}), and so gives no contribution to the 
effective $\hat q$ in the $T\to\infty$ limit.

\subsection{Spacelike Wilson loop}

We now turn to the spacelike prescription for calculating the Wilson loop. 
We will show that a similar qualitative behavior to that of the Euclidean 
path integral shown in (\ref{euc2}) and (\ref{euc3}) also holds for 
spacelike strings.  In particular, the leading contribution is dominated 
by a confining-like ($L$) behavior with no jet quenching-like ($L^2$) 
subleading term, and only an exponentially suppressed longer-string 
contribution has a leading jet quenching-like behavior.  The analogous
results are recorded in (\ref{spacelikeshort}) and (\ref{spacelikelong}),
below.

Since $-G<0$ for spacelike worldsheets, the Nambu-Goto action is imaginary 
and so $\exp\{i S\} = \exp\{\pm A\}$, where $A$ is the positive real area 
of the string worldsheet.  The sign ambiguity comes from the square root in
the Nambu-Goto action.  For our stationary string solutions, the worldsheet 
area is the time of propagation $T$ times the length of the string.  Thus,
with the choice of the plus sign in the exponent, the longest string length
exponentially dominates the path integral, while for the minus sign, the 
shortest string length dominates.  Only the minus sign is physically sensible, 
though, since we have seen in section 3 that the length of the spacelike
string solutions is unbounded from above (since there are solutions with
arbitrarily many turns).  Thus we must pick the minus sign, and, as in the 
Euclidean case, the solution with shortest string length exponentially 
dominates the path integral.

\begin{figure}[h]
\begin{center}
$\begin{array}{c@{\hspace{.75in}}c}
\epsfxsize=1.75in \epsffile{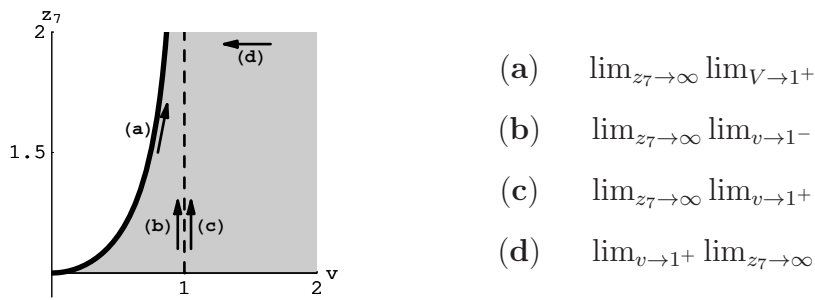} &
\begin{array}{c@{\hspace{.25in}}c} \\ [-1.75in]
{\bf (a)} & \lim_{z_7\to\infty}\lim_{V\to1^+} \\ [.3cm]
{\bf (b)} & \lim_{z_7\to\infty} \lim_{v\to1^-} \\ [.3cm]
{\bf (c)} & \lim_{z_7\to\infty} \lim_{v\to1^+} \\ [.3cm]
{\bf (d)} & \lim_{v\to1^+} \lim_{z_7\to\infty} 
\end{array} \\ [-.3cm]
\end{array}$
\end{center}
\caption[FIG. \arabic{figure}.]{\footnotesize{The shaded
   region is the set of $(v,z_7)$ for which the string worldsheet
   is spacelike and outside the horizon.  The curved boundary
   corresponds to lightlike worldsheets.  The various approaches to the
   lightlike $z_7=\infty$ limit discussed in the text are shown.}}
   \label{fig5}
\end{figure}

As we illustrated in our discussion of the Euclidean Wilson loop, the 
physically sensible limit is to take the quarks infinitely massive 
($z_7\to\infty$) at fixed quark separation $L$.  In the spacelike case,
however, there is a new subtlety: {\it a priori}\/ it is not obvious that
the lightlike limit $V\to1$ will
commute with the $z_7\to\infty$ limit.  Since  
$V= v (1-z_7^{-4})^{-1/2}$, the lightlike limit is $v\to1$ when 
$z_7\to\infty$.  We will examine four different approaches to this
limit, shown in figure \ref{fig5}.\footnote{See \cite{cgg0607}
for a related discussion of the lightlike limit.}
 
\subsubsection*{Limit (a):\ lim$\bf _{z_7\to\infty}$\,lim$\bf_{V\to1^+}$.} 

This is the limit in which we take the lightlike limit at fixed $z_7$, 
then take the mass to infinity.  Recall from (\ref{spacetime}) that a 
spacelike worldsheet requires either $v\ge1$ ($\g^2<0$) for any $z_7$, 
or $v<1$ ($\g^2>1$) and $z_7 <\sqrt\g$.  Since, at fixed $z_7$, $V=1$ 
corresponds to $\g=z_7^2$, we necessarily have $v<1$. Thus, only the 
$v<1$ spacelike solutions discussed in section 3.1.1 will contribute.  
Recall that for these solutions $1\le z^4\le \g^2(1-\a^2)$, where the 
integration constant is in the range $0<\a<v$.  In particular, we must 
keep $\g^2(1-\a^2)\ge z_7^4$ and $\g^2\ge z_7^4$ while taking the $\g^2
\to z_7^4$ limit.  This implies that we must take solutions with $\a\to0$.
However, such solutions necessarily have $L\to0$ (see figure \ref{fig1}),
which contradicts our prescription of keeping $L$ fixed.  Therefore, this 
limit is not interesting.

\subsubsection*{Limit (b):\ lim$\bf_{z_7\to\infty}$\,lim$\bf_{v\to1^-}$.}

Another approach to the lightlike limit takes $v\to1$ from below and then 
takes $z_7\to\infty$.  Then the conditions for a spacelike worldsheet are
automatically satisfied.  Again, only the $v<1$ spacelike solutions 
discussed in section 3.1.1 contribute, but now the $\g^2(1-\a^2)\ge z_7^4$
condition places no restrictions on $\a$.  In particular, this limit will 
exist at fixed $L$.  The behavior of figure \ref{fig1} as $v\to1$ suggests
that, at any given (small) $L$, all the series of string solutions 
illustrated in figure \ref{fig2} occur.  The lengths of these strings 
follow the pattern plotted in figure \ref{fig3}.  Actually, the analysis 
given in appendix B shows that the short (c)$_0$ ``up" string does not exist 
in the limit with fixed $L$.  Thus the long (c)$_0$ ``up" string dominates 
the path integral, with the (a)$_1$ ``down" string and all longer strings 
relatively exponentially suppressed.  

The result from appendix B.1 for the action of the (c)$_0$ string
as a function of $L$ is
\be\label{spacelikeshort}
{\b\hat S\over T\sqrt\l} = -1.31 + {\pi\over2}\, {L\over\b} .
\ee
This result is exact, in the sense that no higher powers of
$L$ enter.  The constant term is from the straight string
subtraction.  The linear term is consistent with energy loss by elastic scattering.

For comparison, the next shortest string is the (a)$_1$
down string solution.  Appendix A computes its action to
be
\be\label{spacelikelong}
{\b \hat S_{\rm long}\over T\sqrt\l} = 
0.941\, {L^2\over\b^2} + {\cal O}(L^4) ,
\ee
which shows the jet-quenching behavior found in \cite{lrw0605}. However, since the contribution from this configuration to the path integral is exponentially suppressed, the actual jet quenching parameter is zero.

\subsubsection*{Limit (c):\ lim$\bf _{z_7\to\infty}$\,lim$\bf_{v\to1^+}$.}

When $v>1$, the string worldsheet is spacelike regardless of the value of 
$z_7$.  Thus, we are free to take the order of limits in many ways.  Limit 
(c) takes $v\to1$ from above at fixed $z_7$ and then takes $z_7\to\infty$.  
We saw in section 3.1.2 that there are always two string solutions for $v>1$: 
a short one with $\a>v$, which turns at $z=z_t:=\g^2(1-\a^2)$, and a long 
one with $\a<v$, which turns at the horizon $z=1$.  Appendix B.2 shows
that in the (c) limit, the short string gives precisely the same contribution
as the (c)$_0$ up string did in the (b) limit.  Similarly, the long string
contirbution coincides with the (a)$_1$ down string.  This agreement is
reassuring, showing that the path integral does not jump discontinuously
between the (b) and (c) limits even though they are evaluated on qualitatively
different string configurations.  (The (b) and (c) limits approach the
lightlike limit in the same way, see figure \ref{fig5}.)

\subsubsection*{Limit (d):\ lim$\bf_{v\to1+}$\,lim$\bf_{z_7\to\infty}$.}  

Limit (d) approaches the lightlike limit in the opposite order to the (c)
limit.  Somewhat unexpectedly, the results for the string action in the
(d) limit are numerically the same as those found in the (b) and (c)
limits.  This is unexpected since the details of evaluating the integrals
in the (c) and (d) limits are substantially different.  We take this 
agreement as evidence that the result is independent of how the lightlike
limit is taken.  (Note that there are, in principle, many different
lightlike limits intermediate between the (c) and (d) limts.)

\appendix

\section{Euclidean action}

\subsubsection*{Euclidean string solutions}

Wick rotate $x_0 \to i x_4$ in (\ref{metric}), and adopt the rotated 
boundary conditions (\ref{bcs}) with $x_0\to x_4$.  Then the Euclidean 
version of the $[v_\perp]$ embedding (\ref{shape}) becomes
\be\label{eucshape}
x_4=\tau,\qquad x_1=v\tau, \qquad x_2=\sigma, \qquad x_3=0,
\qquad r=r(\sigma) .
\ee
One then finds \cite{aev0608} that the integral expressions (\ref{L}) and 
(\ref{S}) for the quark separation and string action stay the same except 
for the replacement
\be
\g^2\to\g_E^2:={1\over1+v^2}.
\ee

Real Euclidean string configurations must have the integration constant 
$a^2$ be positive, to have positive $G$.  Then an analysis of the Euclidean 
string equations of motion \cite{aev0608} shows that real solutions can 
exist for any $v$ as long as the string is at radii satisfying
\be\label{eradii} 
z^4\ >\ \mbox{max}\left\{1,\ \g_E^2(1+a^2)\right\}.
\ee
(These are for the string configurations with endpoint ``velocity" 
perpendicular to their separation.)

For $a>v$ and $v$ sufficiently large, there is a unique Euclidean solution 
with turning point at $z^4 = z_t^4 :=(1+a^2)/(1+v^2)>1$.  We call these the 
``short string" solutions.  For $a<v$ there is a branch of Euclidean 
solutions which have the radial turning point on the black hole horizon 
$z=1$.  These are the ``long string" solutions.  The solution with the 
smallest energy dominates the path integral.  The energy of the Euclidean 
string configurations is given by $E=S/T$, where $S$ is the Nambu-Goto action 
and $T$ is the time interval.  For $L$ less than a critical value,
the energetically favorable state is the short string \cite{aev0608}.  

The Euclidean rotation of strings whose endpoints have lightlike worldlines 
are those with Euclidean worldsheet (\ref{eucshape}) with $V=1$.  By 
(\ref{realV}) this is when $v = \sqrt{1-z_7^{-4}}$.  But since nothing 
special happens to the Euclidean string configurations at this velocity, 
we will do our computations below at arbitrary $v$, and specialize to the 
lightlike value at the end.

We are interested in evaluating the action for this string in the small
$L$  and small $\e:=z_7^{-1}$ (large mass) limit.  These two limits do 
not commute, so we evaluate them separately in the two different orders.

\subsubsection*{$L\to0$ at fixed (small) $\e$}

{}From (\ref{L}) with $\g\to\g_E$, the $L\to0$ limit corresponds to taking 
$z_t\to z_7$.  So introduce a small parameter $\d$ defined by
\be
z_t^4 := {z_7^4 \over (1+\d)^4} = {1\over \e^4 (1+\d)^4}.
\ee
Thus $\delta$ replaces the parameter $a$.

Changing variables to $y=\e(1+\d) z$, (\ref{L}) and (\ref{S}) can be 
rewritten in terms of $\d$ as
\bea\label{edexp}
{L\over\b} &=&
{2\over\pi} \e (1+\d)
\int_1^{1+\d}\!\!\! {dy\, \sqrt{1-\g_E^2\e^4(1+\d)^4}
\over\sqrt{(y^4-1)(y^4-\e^4(1+\d)^4)}},
\nonumber\\
{\b S\over T \sqrt\l} &=&
{1\over\g_E\e(1+\d)}
\int_1^{1+\d}\!\!\! {dy\,\left[y^4-\g_E^2\e^4(1+\d)^4\right]
\over\sqrt{(y^4-1)(y^4-\e^4(1+\d)^4)}} .
\eea
Systematically expanding in small $\delta$ gives series expressions in 
terms of integrals of the form 
\be\label{jnm}
J_{nm} :=\int_1^{1+\d} \frac{y^{4m}dy}{(y^4-1)^{1\over2} 
(y^4-\e^4)^{{1\over2}+n}},
\ee
which have a series expansion of the form $\sum_{n=0}^\infty c_n
\d^{n+{1\over2}}$, but whose coefficients $c_n(\e)$ lack closed-form 
expressions.  Nevertheless, the $J_{mn}$ are uniformly convergent for 
$\e<1$, so we can expand the integrands in power series in small $\e$ 
to find
\bea
{L\over\b} &=& {2\e\d^{1/2}\over\pi} 
\Bigl\{ \left[1 + \textstyle{1\over2}(1-\g_E^2)\e^4 
+ {\cal O}(\e^8) \right] \nonumber\\
&& \qquad \qquad \mbox{} + \d \left[ \textstyle{1\over12} 
+ {1\over24}(33-49\g_E^2)\e^4 + {\cal O}(\e^8) \right] 
+ {\cal O}(\d^2) \Bigr\},
\nonumber\\
{\b S\over T\sqrt\l} &=& {\d^{1/2}\over\g_E\e}
\Bigl\{ \left[1 + \textstyle{1\over2}(1-\g_E^2)\e^4 
+ {\cal O}(\e^8) \right] \nonumber\\
&& \qquad\qquad \mbox{} + \d \left[ -\textstyle{5\over4} 
+ {17\over24}(1-\g_E^2)\e^4 + {\cal O}(\e^8) \right] 
+{\cal O}(\d^2) \Bigr\}. \nonumber
\eea
Eliminating $\d$ between these two expressions order-by-order in $\d$ then 
gives the action as a function of $L$ and $\e$:
\be\label{e0res}
{\b S\over T\sqrt\l}= 
{\pi\over2\g_E}{L\over\b} \left[{1\over\e^2} +{\cal O}(\e^6)\right]
- {\pi^3\over6\g_E}{L^3\over\b^3} \left[{1\over\e^4}
-{1\over2}(2-\g_E^2)  +{\cal O}(\e^4)\right]
+{\cal O}(L^5).
\ee
The lightlike limit corresponds to taking $\g_E=(2+\e^4)^{-1/2}$, giving
\be\label{e1res}
{\b S\over T\sqrt\l}= {\pi\over\sqrt2}{L\over\b} 
\left[{1\over\e^2}+{1\over4}\e^2+{\cal O}(\e^6)\right]
-{\pi^3\over3\sqrt2}{L^3\over\b^3} \left[{1\over\e^4}
-{1\over2}+{\cal O}(\e^4)\right]+{\cal O}(L^5).
\ee

Note that because of the nice convergence properties of the integrals in 
(\ref{jnm}), the order of limits as $\e\to0$ and $\d\to0$ does not affect 
this result.  The limiting case where $\e\to0$ with $\d$ fixed (and small) 
corresponds to taking $L\propto\e$.  Thus the result (\ref{e1res}) is valid 
for all limits $\e\to0$ with $L\to0$ as $L\propto\e$ or faster.

\subsubsection*{$\e\to0$ limit at fixed $L$}

To keep $L$ fixed, examination of (\ref{edexp}) shows that we need to scale 
$\d\to\infty$ as $\e\to0$ keeping $\e(1+\d)$ fixed.  So change variables in
(\ref{edexp}) from $\d$ to $\ell := \e(1+\d)$.  Since for fixed $\ell<1$ the 
integral is convergent, we can take the $\e\to0$ limit directly, and then 
expand in powers of $\ell$ to get
\be\label{lexp}
{L\over\b} =
{2\over\pi} \ell
\int_1^{\infty}\!\!\! {dy\, \sqrt{1-\g^2_E\ell^4}
\over\sqrt{(y^4-1)(y^4-\ell^4)}}
= {2\over\sqrt\pi} {\Gtf\over\Gof}\,\ell
\left(1+{3-5\g^2_E\over10}\ell^4+{\cal O}(\ell^8)\right).
\ee
Similarly, the $S$ integral is
\bea
{\b S\over T \sqrt\l} &=& \lim_{\e\to0}
{1\over\g_E\ell}
\int_1^{\ell/\e} {dy \left[y^4-\g^2_E\ell^4\right]
\over\sqrt{(y^4-1)(y^4-\ell^4)}} 
\nonumber\\
&=&  {1\over\e\g_E}
-{\sqrt\pi\over\ell\g_E}  
{\Gtf\over\Gof}
\left\{ 1-{1-2\g^2_E\over2}\ell^4+{\cal O}(\ell^8) \right\},
\eea
where we used the fact that only the leading term at small $\ell$ diverges 
as $1/\e$, and so the $\e\to0$ limit can be taken directly in all the other 
terms.

Since $S$ is divergent as $\e\to0$, we regulate the action by subtracting 
the action of a pair of straight strings with the same boundary conditions.
(See, however, \cite{cgg0607} for a discussion of an alternative 
regularization procedure.)  The straight string solutions have embeddings
\be
x_4=\tau,\qquad x_1=v\tau,\qquad x_2=\pm L/2,\qquad
x_3=0,\qquad r=\sigma,
\ee
with boundary conditions $0\le\tau\le T$ and $r_0\le\sigma\le r_7$.  The 
action evaluated on these two solutions is then
\be
{\b S_0\over T\sqrt\l} = \lim_{\e\to0}
{1\over\g_E} \int_1^{1/\e} dz\,\sqrt{z^4-\g_E^2\over z^4-1}
\nonumber\\
= {1\over\e\g_E} - {\sqrt\pi\over\g_E} {\Gtf\over\Gof}
{}_2F_1[\textstyle -{1\over2}, -{1\over4}, {1\over4}, \g_E^2] ,
\ee
where ${}_2F_1$ is a hypergeometric function.  Thus the regularized action 
$\hat S := S-S_0$ is
\be\label{sexp}
{\b \hat S\over T \sqrt\l} =
{\sqrt\pi\over\g_E} {\Gtf\over\Gof}
\left\{
{}_2F_1[{\textstyle -{1\over2}, -{1\over4}, {1\over4}}, \g_E^2]
-{1\over\ell}+{1-2\g^2_E\over2}\ell^3+{\cal O}(\ell^7) \right\}.
\ee

Eliminating $\ell$ order-by-order between (\ref{lexp}) and (\ref{sexp}) gives
\be
{\b \hat S\over T \sqrt\l} = 
-{\Gtf^4\over\pi^2\g_E}{\b\over L}
+{\Gtf^2\over\sqrt{2\pi}\g_E}
{}_2F_1[{\textstyle -{1\over2}, -{1\over4}, {1\over4}}, \g_E^2]
+\Gof^4 (2-5\g_E^2) {L^3\over\b^3}
+{\cal O}(L^7) .
\ee
The lightlike limit corresponds to $\g_E^2=1/2$, giving
\be\label{e2res}
{\b \hat S\over T \sqrt\l} = 
-{0.32}\,{\b\over L}+1.08 
-{0.76}\,{L^3\over\b^3}
+{\cal O}(L^7) .
\ee

\subsubsection*{$\e\to0$ limit at fixed $L$: long string}

For comparison purposes, we also compute the contribution to the Wilson loop 
from the long Euclidean string solution.  This is the solution with turning 
point at the horizon, $z_t=1$.  The integral expression for $L$ is convergent 
as $\e\to\infty$ and, to keep $L$ fixed and small, we just need to keep $a$ 
fixed and small.  So expanding in small $a$ gives
\bea
{L\over\b} &=& {2a\g_E\over\pi} \int_1^{\infty} 
{dz\over\sqrt{(z^4-1)(z^4-\g_E^2[1+a^2])}}, \\
&=& a\g_E\,2^{3/2}\pi^{1/2}\Gof^2
\Bigl[ {}_2F_1[{\textstyle {3\over4}, {3\over2}, {5\over4}}, \g_E^2] 
- {\textstyle{3\over5}}\,\g_E^2\,
{}_2F_1[{\textstyle {3\over2}, {7\over4}, {9\over4}}, \g_E^2] 
\nonumber\\
&&\qquad\qquad\qquad\qquad\qquad \mbox{}+ 
{\textstyle{3\over10}} a^2 \g_E^2\, 
{}_2F_1[{\textstyle {3\over2}, {7\over4}, {9\over4}}, \g_E^2]
+{\cal O}(a^4) \Bigr]. \nonumber
\eea 
The same expansion of the regularized action gives
\bea
{\b \hat S_{\rm long}\over T\sqrt\l} &=& {1\over\g_E}
\int_1^{\infty}{dz\,\sqrt{z^4-\g_E^2}\over\sqrt{z^4-1}}
\left( {\sqrt{z^4-\g_E^2}\over
\sqrt{z^4-\g_E^2[1+a^2]}} 
- 1 \right) \\
&=& a^2\g_E\,2^{-1/2}\pi^{3/2} \Gof^{-2}
\Bigl[ {}_2F_1[{\textstyle {3\over4}, {3\over2}, {5\over4}}, \g_E^2] 
- {\textstyle{3\over5}}\,\g_E^2\,  
{}_2F_1[{\textstyle {3\over2}, {7\over4}, {9\over4}}, \g_E^2] 
\nonumber\\
&&\qquad\qquad\qquad\qquad\qquad \mbox{}+ 
{\textstyle{9\over20}} a^2 \g_E^2\, 
{}_2F_1[{\textstyle {3\over2}, {7\over4}, {9\over4}}, \g_E^2]
+{\cal O}(a^4) \Bigr]. \nonumber
\eea
Eliminating $a$ order-by-order between these two expressions gives
\be
{\b \hat S_{\rm long}\over T\sqrt\l} = {L^2\over\b^2}\, 
{5\sqrt\pi\over8\sqrt2\g_E} \Gof^2
\left(5\,{}_2F_1[{\textstyle{3\over2},{3\over4},{5\over4}},\g_E^2] - 
3\,\g_E^2\,{}_2F_1[{\textstyle{3\over2},{7\over4},{9\over4}},\g_E^2]
\right)^{-1} +{\cal O}(L^4) .
\ee
The lightlike limit is $\g_E^2=1/2$, giving
\be
{\b \hat S_{\rm long}\over T\sqrt\l} = +2.39 \, {L^2\over\b^2} 
+{\cal O}(L^4) .
\ee

\section{Spacelike action}

Here we calculate the regulated action of spacelike strings in the various 
limits described in section 4, keeping $L$ fixed and small.  Specifically, 
this requires expanding the quark separation (\ref{L}) and string action 
(\ref{S}) in terms of the appropriate small parameter, and then eliminating 
that parameter to obtain $\hat S$ as a function of $L$.  Recall from the 
discussion in section 3 that for spacelike strings the integration constant 
$a$ is imaginary, so we replace $a^2 = -\a^2$ with $\a^2>0$.  Also, recall
that $\g^2:=(1-v^2)^{-1}$ and $\e:=z_7^{-1}$.  Then our main equations 
(\ref{L}) and (\ref{S}) become
\be\label{Lprime}
{L\over\b}=\left|\frac{2\a\g}{\pi} \int_{z_t}^{1/\e} 
{dz\over\sqrt{(z^4-1)(z^4-\g^2(1-\a^2))}}\right|, 
\ee
and
\be\label{Sprime}
{\b\hat S\over T\sqrt\l}= {1\over|\g|}
\left\{
\left|\int_{z_t}^{1/\e} {(z^4-\g^2)\, dz \over
\sqrt{(z^4-1)(z^4-\g^2(1-\a^2))}} \right|
- \left|\int_1^{1/\e} dz\,\sqrt{z^4-\g^2\over z^4-1}
\right|\right\},
\ee
where $z_t$ is either $z_t=1$ or $z_t^4 := \g^2(1-\a^2)$, depending on which
configuration we are considering, and we have subtracted the action of two 
straight strings to regulate the action.   These expressions are valid for 
string solutions which have a single turning point, which will be the only 
ones we evaluate.

\subsection{The (b) limit}

\subsubsection*{The (c)$_0$ ``up string" solutions.}

The (c)$_0$ string solutions have a turning point at $z_t^4=\g^2(1-\a^2)>z_7^4$.
Since in this limit we first take $v\to1^-$ ($\g\to+\infty$) before taking 
$\e\to0$, (\ref{Lprime}) becomes
\be\label{Lc0}
{L\over\b}=\frac{2\a\g}{\pi} \int^{\g^{1/2}(1-\a^2)^{1/4}}_{1/\e} 
{dz\over\sqrt{(z^4-1)(\g^2(1-\a^2)-z^4)}}. 
\ee
The upper limit of this integral gives a contribution that scales as $\a
(1-\a^2)^{-3/4}\g^{-1/2}$, while the lower limit gives $\a(1-\a^2)^{-1/2}\e$.
Therefore, to keep $L$ fixed in this limit requires that either the 
contribution from the upper limit or from the lower limit remains finite 
and non-zero.  In order for the upper limit to remain finite and non-zero, it is required $1-\a^2\sim\g^{-2/3}$. However, then the lower limit contributes 
$\sim\g^{1/3}\e$ which diverges as we take $\g\to\infty$.  Thus, this scaling 
does not keep $L$ finite.  If, instead, we demand that the lower limit remain 
finite, we must take $1-\a^2\sim \e^2$.  This then implies that the upper 
limit contributes $\sim(\g\e^3)^{-1/2}$ which vanishes in the $\g\to\infty$ 
limit, and so $L$ remains finite.

We have thus found that there is a single (b) limit of the (c)$_0$ up strings 
which keeps the quark separation finite.  This limit keeps 
\be\label{c0limit}
\d^2 := {\e^2\over1-\a^2}
\ee
fixed and gives $L\sim \d$ for small $\d$.  Since for fixed $L$ and $\e$ and 
$\g\to\infty$ this limit keeps $\a$ fixed away from $\a=1$, then from the 
discussion in section 3.1.1 we see that this solution corresponds to the long
(c)$_0$ string (see figure \ref{fig1}).  In particular, the short (c)$_0$ 
string does not contribute.

Plugging (\ref{c0limit}) into (\ref{Lc0}), changing variables to $y = 
(\g\e/\d)^{-1/2} z$, and expanding the $(z^4-1)^{-1/2}$ factor for large $z$
gives a series of hypergeometric integrals which are finite in the 
$\g\to\infty$ limit, giving
\be\label{slc0L}
{L\over\b} 
= \lim_{\e\to0} {2\over\pi} \sqrt{\d^2-\e^2} 
\left[1 + {\textstyle{1\over5}}\e^4 + {\cal O}(\e^8) \right] 
= {2\over\pi}\d .
\ee

Similarly, the regularized action (\ref{Sprime}) becomes
\be\label{Sc0}
{\b\hat S\over T\sqrt\l}= {1\over\g}
\left\{
\int_{1/\e}^{\sqrt{\g\e/\d}}\!\!\! {(\g^2-z^4)\, dz \over
\sqrt{(z^4-1)(\g^2\e^2\d^{-2}-z^4)}} 
- \int_1^{1/\e} dz\,\sqrt{\g^2-z^4\over z^4-1}
\right\},
\ee
which, upon making the same change of variables and taking the large $\g$
limit, becomes
\be\label{slc0S}
{\b\hat S\over T\sqrt\l}
= - {\sqrt\pi\over4}{\Gof
\over\Gtf}
+\lim_{\e\to0}\, (\d+\e) \left[1+ {\textstyle{1\over10}} \e^4 
+{\cal O}(\e^8)\right] 
= -1.31 + \d.
\ee
Eliminating $\d$ between (\ref{slc0L}) and (\ref{slc0S}) gives
\be\label{blimc0}
{\b\hat S\over T\sqrt\l} = -1.31 + {\pi\over2}\, {L\over\b} .
\ee

\subsubsection*{The (a)$_1$ ``down string" solution.}

The (a)$_1$ string descends from the D7-brane and turns at the horizon, so 
that the quark separation (\ref{Lprime}) is given by
\be\label{La1}
{L\over\b}={2\a\g\over\pi} \int_1^{1/\e} 
{dz\over\sqrt{(z^4-1)(\g^2(1-\a^2)-z^4)}}.
\ee
In the $\g\to\infty$ limit, $L$ is kept finite for finite $\a$.  Taking the 
limit directly gives
\be\label{sla1L}
{L\over\b}
= {1\over2\sqrt\pi}{\Gof
\over\Gtf}{\a\over\sqrt{1-\a^2}} .
\ee
Similarly, the limit of the regularized action gives
\be\label{Sa1}
{\b\hat S_{\rm long}\over T\sqrt\l}
={\sqrt\pi\over4}{\Gof
\over\Gtf}
\left({1\over\sqrt{1-\a^2}}-1\right).
\ee
Eliminating $\a$ between these two expressions and expanding in small $L$ 
yields
\be\label{blima1}
{\b\hat S_{\rm long}\over T\sqrt\l}
= {\pi^{3/2}\over2}{\Gtf
\over\Gof}
{L^2\over\b^2} + {\cal O}(L^4) 
= 0.941\, {L^2\over\b^2} + {\cal O}(L^4) .
\ee

\subsection{The (c) limit.}

The (c) and (d) limits take $v\to1$ with $v>1$.  In this range it is
convenient to define
\be
\tg^2 := -\g^2 = {1\over v^2-1},
\ee
so the $v\to1^+$ limit takes $\tg^2\to+\infty$.  The spacelike string 
solutions for $v>1$ were discussed in section 3.1.2, where we found that 
there are two solutions: a short string with turning point $z_t^4 =
\tg^2(\a^2-1)$ and a long string with turning point at the horizon
$z_t=1$.

\subsubsection*{The short string solution.}

For this solution, the integral expression for the quark separation
(\ref{Lprime}) takes the form 
\be\label{climLs}
{L\over\b}=
{2\over\pi}\a\tg\int_{\tg^{1/2}(\a^2-1)^{1/4}}^{1/\e} 
{dz\over\sqrt{(z^4-1)(z^4-\tg^2(\a^2-1))}}.
\ee
The (c) limit takes $\tg\to\infty$ first before taking $\e\to0$.
Examination of (\ref{climLs}) shows that, in this limit, $L$ remains
finite if one takes $\a\to1^+$ in such a way that
\be
\d^2 := \e^6 \tg^2[ 1 - \e^4 \tg^2 (\a^2-1)]
\ee
remains fixed.  Eliminating $\a$ in favor of $\d$ in (\ref{climLs})
and changing variables to $y=\e z$ gives
\be
{L\over\b}=
{2\over\pi}\sqrt{1-\tg^{-2}\d^2\e^{-6}+\e^4\tg^2} 
\int_{(1-\tg^{-2} \d^2\e^{-6})^{1/4}}^{1} 
{\e\,(1-\e^4y^{-4})^{-1/2}\, dy\over y^2\sqrt{y^4-1+\tg^{-2}\d^2\e^{-6}}}.
\ee
For small fixed $\e$, expanding the numerator of the integrand
in a power series, performing the integrals, and taking the $\tg\to\infty$
limit yields
\be\label{climLs2}
{L\over\b} = \lim_{\e\to0}{\d \over\pi}
\left[1+{\cal O}(\e^4)\right]= {\d\over\pi}.
\ee
Similarly, in the same limit, the regularized action
\be
{\b\hat S\over T\sqrt\l}= \int_{\tg^{1/2}(\a^2-1)^{1/4}}^{1/\e} 
{\tg^{-1} (z^4+\tg^2)\, dz \over \sqrt{(z^4-1)(z^4-\tg^2(\a^2-1))}}
- \int_1^{1/\e} {dz\over\tg} \sqrt{z^4+\tg^2\over z^4-1} ,
\ee
becomes
\be\label{climSs2}
{\b\hat S\over T\sqrt\l} = 
\lim_{\e\to0}\left[{\d \over2}
-{\sqrt\pi\over4}{\Gof\over\Gtf}+{\cal O}(\e)\right]= {\d\over2}-1.31.
\ee
Eliminating $\d$ between (\ref{climLs2}) and (\ref{climSs2}) gives
\be\label{climSs3}
{\b\hat S\over T\sqrt\l} = 
-1.31+{\pi\over2} {L\over\b} .
\ee

\subsubsection*{The long string solution.}

The turning point for the long string is at the horizon, so 
\be\label{climLl}
{L\over\b}={2\a\tg\over\pi} \int_{1}^{\e^{-1}} 
{dz\over\sqrt{(z^4-1)(z^4+\tg^2(1-\a^2))}}, 
\ee
from which it follows that $L$ is kept finite in the $\tg\to\infty$ limit for finite $\a$.  Taking the limit directly then gives
\be\label{climL2}
{L\over\b} = {1\over2\sqrt\pi}{\Gof\over\Gtf}
{\a\over\sqrt{|1-\a^2|}}.
\ee
Similarly, the limit of the regularized action gives
\be\label{climS1}
{\b\hat S_{\rm long}\over T\sqrt\l}=
{\sqrt\pi\over4}{\Gof\over\Gtf}
\left({1\over\sqrt{|1-\a^2|}}-1\right).
\ee
Eliminating $\a$ between (\ref{climL2}) and (\ref{climS1}) and expanding in
powers of small $L$ gives
\be\label{climS2}
{\b\hat S_{\rm long}\over T\sqrt\l}
= {\pi^{3/2}\over2}{\Gtf\over\Gof}
{L^2\over\b^2} + {\cal O}(L^4) 
= 0.941\, {L^2\over\b^2} + {\cal O}(L^4) .
\ee
Note that this calculation is essentially identical to that of the (a)$_1$
string in the (b) limit.

\subsection{The (d) limit.}

\subsubsection*{The short string solution.} 

The (d) limit takes $\e\to0$ first, then $\tg\to\infty$.  Examination of 
(\ref{climLs}) shows that, in this limit, $L$ remains finite if one takes 
$\a\to1^+$ in such a way that
\be
\d := \tg^{-1/2} (\a^2-1)^{-3/4}
\ee
remains fixed.  Eliminating $\a$ in favor of $\d$ in (\ref{climLs})
and changing variables to $y=\tg^{-1/3}\d^{1/3} z$, the $\e$ and $\tg$
limits can be taken directly to give
\be\label{dlimLs1}
{L\over\b}
={2\over\pi}\d\int_1^{\infty} {y^{-2} dy\over\sqrt{y^4-1}}
={2\over\sqrt\pi}{\Gtf\over\Gof}\, \d.
\ee
The regularized action can be written as
\be\label{dlimSs1}
{\b\hat S\over T\sqrt\l}
= \int_{(\tg/\d)^{1/3}}^{1/\e} 
{\tg^{-1} (z^4+\tg^2)\, dz \over \sqrt{(z^4-1)(z^4-(\tg/\d)^{4/3})}}
- \int_1^{1/\e} {dz\over\tg} \sqrt{z^4+\tg^2\over z^4-1} .
\ee
To evaluate this expression as $\tg\to\infty$ (after $\e\to0$), split the 
ranges of integration into $z<\tg^{1/2}$ and $z>\tg^{1/2}$.  After the pieces 
of the integrals for $z>\tg^{1/2}$ which are divergent at $\e\to0$ are 
canceled, the remainder is easily seen to vanish in the $\tg\to\infty$ 
limit.  The integrals for $z<\tg^{1/2}$ are evaluated to give
\be\label{dlimSs2}
{\b\hat S\over T\sqrt\l}
=-{\Gof^2\over4\sqrt{2\pi}} + {\Gtf^2\over\sqrt{2\pi}}\,\d 
\ee
in the $\tg\to\infty$ limit.  Eliminating $\d$ between (\ref{dlimLs1}) and 
(\ref{dlimSs2}) gives
\be\label{dlimSs3}
{\b\hat S\over T\sqrt\l} = 
- {\Gof^2\over4\sqrt{2\pi}} + {\pi\over2}{L\over\b}
= -1.31 + {\pi\over2}\, {L\over\b} .
\ee

\subsubsection*{The long string solution.}

For the long strings with $v>1$, recall that $\a<1$ and the
turning point is  at the horizon, so that
\be\label{dlimLl1}
{L\over\b}=\frac{2\a\tg}{\pi} \int_{1}^{\infty} 
{dz\over\sqrt{(z^4-1)(z^4+\tg^2(1-\a^2))}}, 
\ee
where we have already taken the $\e\to0$ limit since
the integral is convergent.  $L$ is kept small and finite as
$\tg\to\infty$ if $\a$ is kept small and fixed.  Then the
above integral can be evaluated by splitting the range of
integration into $z^4 > \tg^2(1-\a^2)$ and $z^4<\tg^2(1-\a^2)$.
The upper range is easily seen to give a vanishing contribution
in the $\tg\to\infty$ limit, while the lower range gives
\be\label{dlimLl2}
{L\over\b}
=\lim_{\tg\to\infty}{2\a\over\pi} \int_{1}^{\g^{1/2}} 
{dz(1+z^4\tg^{-2}(1-\a^2)^{-1})^{-1/2}\over\sqrt{1-\a^2}\sqrt{z^4-1}} 
= {1\over2\sqrt\pi}{\Gof\over\Gtf}\,{\a\over\sqrt{1-\a^2}} .
\ee
Similarly evaluating the integral for the action gives
\be\label{dlimSl1}
{\b\hat S_{\rm long}\over T\sqrt\l}
={\sqrt\pi\over4}{\Gof\over\Gtf}\left({1\over\sqrt{1-\a^2}}-1\right).
\ee
Eliminating $\a$ between (\ref{dlimLl2}) and (\ref{dlimSl1}) and
expanding in powers of $L$ gives
\be\label{dlimSl2}
{\b\hat S_{\rm long}\over T\sqrt\l} 
= {\pi^{3/2}\over2}{\Gtf\over\Gof}
{L^2\over\b^2} + {\cal O}(L^4) 
= 0.941\, {L^2\over\b^2} + {\cal O}(L^4) .
\ee
Note that this calculation gives the same result as that of the (a)$_1$
(down) string in the (b) limit, and the long string in the (c) limit.

\section*{Acknowledgments}

We would like to thank Daniel Cabrera, Mariano Chernicoff, Paul Esposito, 
Joshua Friess, Antonio Garc\'ia, Richard Gass, Steven Gubser, Alberto 
G\"uijosa, Dan Kabat, Andreas Karch, Che Ming Ko, Hong Liu, Georgios Michalogiorgakis, 
Peter Moomaw, Silviu Pufu, Krishna Rajagopal, Ralf Rapp, Lorenzo Ravagli, 
Kostas Sfetsos, Hendrik van Hees, Urs Wiedemann, Rohana Wijewardhana and 
John Wittig for helpful conversations and correspondence.  
P.C.A. and M.E. are supported by DOE grant FG02-84ER-40153 and J.F.V.P. is supported in part by DOE grant FG03-95ER-40917.

\end{document}